# A fully-digital semi-rotational frequency detection algorithm for bang-bang CDRs

Soon-Won Kwon, Hanho Choi, Younho Jeon, Bongjin Kim, WooHyun Kwon , Homin Park, Kyeongha Kwon, Gain Kim and Hyeon-Min Bae, *Member, IEEE*

*Abstract*—This work presents a new frequency acquisition method using semi-rotational frequency detection (SRFD) algorithm for a reference-less clock and data recovery (CDR) in a serial-link receiver. The proposed SRFD algorithm classifies the bang-bang phase detector(BBPD) outputs to estimate the current phase state, and detects the frequency mismatch between the input data and the sampling clock. The VCO-track path in a digital loop filter (DLF) enables online calibration of a drifted frequency of VCO caused by temperature or voltage variation after a frequency acquisition. The proposed algorithm can be implemented as a digitally-synthesized circuit, lowering design efforts for referenceless CDRs. A 10 Gbps transceiver IC with the proposed algorithm, fabricated in a 65nm CMOS process, demonstrates successful recovery of the input phase without any reference clock.

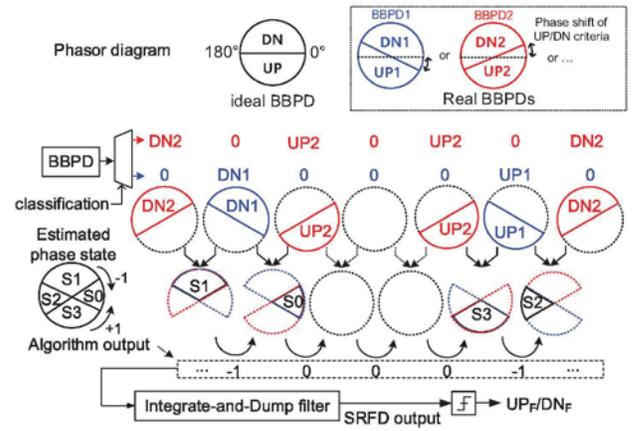

Fig. 1. Basic principle of SRFD algorithm

## I. INTRODUCTION

IN a high-speed serial communication system, various methods for detecting frequency information from input data have been proposed for cost reduction by eliminating an external reference clock source. A majority of those methods rely on flip-flops operating at line-rate for sampling multi-phase clock signals at the edge of input data, leading to design challenges at high-speed application [1]–[12]. A 4x oversampling scheme for frequency detection is also disadvantageous as line-rate 0.25UI-spaced multi-phase clocks are vulnerable to phase mismatch caused by PVT variations [13], [14]. Moreover, additional samplers increase the power consumption and the area occupancy of the receiver. A stochastic referenceless CDR scheme requires line-rate clock dividers and the information of the input signal's transition density, which limits its application flexibility [15]–[19]. Recently, a synthesizable digital algorithm for frequency detection has been reported [20]. However, this algorithm requires a specific baud-rate CDR architecture limiting its application to conventional CDRs. This work proposes a new frequency detection algorithm by modifying the principles of the conventional rotational frequency detection (RFD) algorithm neither using 4x oversampling nor with any additional circuitries. This algorithm can be implemented with standard digital logic synthesis and automatic back-end design flow to reduce design effort. Furthermore, its compatibility with conventional CDR architectures allows the proposed algorithm to be employed in a broad application range in wireline industries.

The rest of the paper is organized as follows. Section II introduces the underlying principles of the proposed algorithm and presents simulation results. The transceiver architecture and the circuit implementation for demonstrating the feasibility of the proposed algorithm are shown in Section III. Section IV presents the measured results, and Section V concludes the paper.

## II. SEMI-ROTATIONAL FREQUENCY DETECTION(SRFD) ALGORITHM

Fig. 1 shows the operation principle of the proposed SRFD algorithm. An ideal BBPD determines UP and DN signals based on 0° and 180° phase of the input data. However, An actual BBPD has slightly different UP/DN criteria for each output for various reasons such as inter-symbol interference(ISI) and voltage offset of the comparator. The proposed algorithm roughly classifies the actual BBPD output into two groups having different UP/DN criteria each other. Those two UP/DN boundaries divide the phase diagram into more than three phase states. One BBPD output cannot specify the phase location of the current input data as one of these phase states. However, assuming that the phase of the input data does not change significantly during the phase detection period, the

Soon-Won Kwon, Hanho Choi, Bongjin Kim, WooHyun Kwon , Homin Park, Kyeongha Kwon, Gain Kim and Hyeon-Min Bae are with the school of Electrical Engineering, Korea Advanced Institute of Science and Technology(KAIST), Daejeon, Korea (e-mail: ksw8538@kaist.ac.kr; zegalho@kaist.ac.kr; bjkim3927@kaist.ac.kr; dngusrnjs@kaist.ac.kr; phm123@kaist.ac.kr; kyeonghaha@kaist.ac.kr; gikim@kaist.ac.kr; hmbae@kaist.ac.kr).

Younho Jeon is with Samsung Electronics, Suwon, Korea (e-mail: younho.jeon@samsung.com).

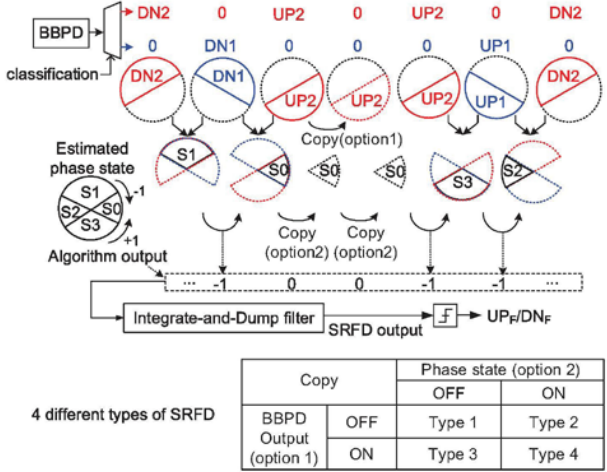

Fig. 2. 4 types of SRFD algorithm

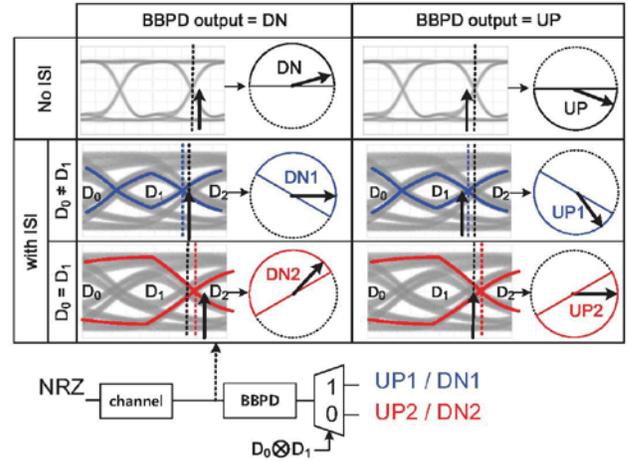

Fig. 3. The classification of BBPD output using inter-symbol interference(ISI)

current phase state can be estimated by combining successive outputs from two different BBPDs. For example, if 'DN2' from the $2^{nd}$ group and 'DN1' from the $1^{st}$ group are output sequentially as shown in Fig. 1, the current phase state can be estimated as 'S1'. Thus, the rotational direction of these estimated states shows the sign of the frequency error between the input data and the sampling clock. The digital algorithm outputs +1, -1, or 0 depending on whether the estimated phase state rotates clockwise or counterclockwise or stays in the same position, respectively. Then, the integrate-and-dump filter accumulates the algorithm outputs to produce a single $UP_F/DN_F$ signal. The proposed frequency detection scheme is named as semi-RFD(SRFD) since it has a similar operation method to conventional RFDs but with a different state estimation principle.

The proposed algorithm can be modified to four different ways depending on the signal processing scheme in dealing with the absence of BBPD output or that of estimated phase state, as shown in Fig.2. These four methods show some performance differences one another, but all of those extract frequency information successfully from the input NRZ signal. To validate the proposed scheme, this paper explores the type 2 SRFD algorithm as an example.

Meanwhile, many different SRFD algorithms are also available depending on the way of classification of the BBPD output into two groups. The SRFD algorithm in this paper exploits the ISI of input data for grouping BBPD output as shown in Fig.3. Due to ISI, the input data creates multiple zero-crossing points at the edge phase, resulting in a data-dependent BBPD output. For example, the zero-crossing point moves forward or backward in time, depending on whether two consecutive bits before the edge are equal or not [21]. Thus, the two groups of BBPD output that are classified by the result of the exclusive-OR of two previous bits have different UP/DN criteria. The SRFD algorithm using this classification scheme is named as dispersion-based SRFD. Previous research [20] also detects frequency information by using ISI of input data but requires multilevel sampling of the input data. On the other hand, the proposed algorithm only requires the conventional BBPD output for its operation, hence it can be easily applied to the conventional CDR architectures widely-employed by industries.

Fig.4 shows MATLAB simulation results of two BBPD output groups when the input NRZ data passes through various channel loss models. As the input signal is more dispersed by ISI, the gain of both BBPD output groups decreases while the difference between their UP/DN criteria increases. Fig.5 shows the integrate-and-dump filter output of type 2 SRFD algorithm under same channel conditions in Fig.4. In this simulation, the edge sampling period is set to 2UI, and the integrate-and-dump filter adds 100000 algorithm outputs to generate one SRFD output. The proposed SRFD algorithm successfully detects the sign of frequency error when there is a loss of about -5dB to -15dB at the Nyquist frequency ($f_{Ny}$). The detection range of the dispersion-based SRFD is about $-22\%$ to $22\%$.

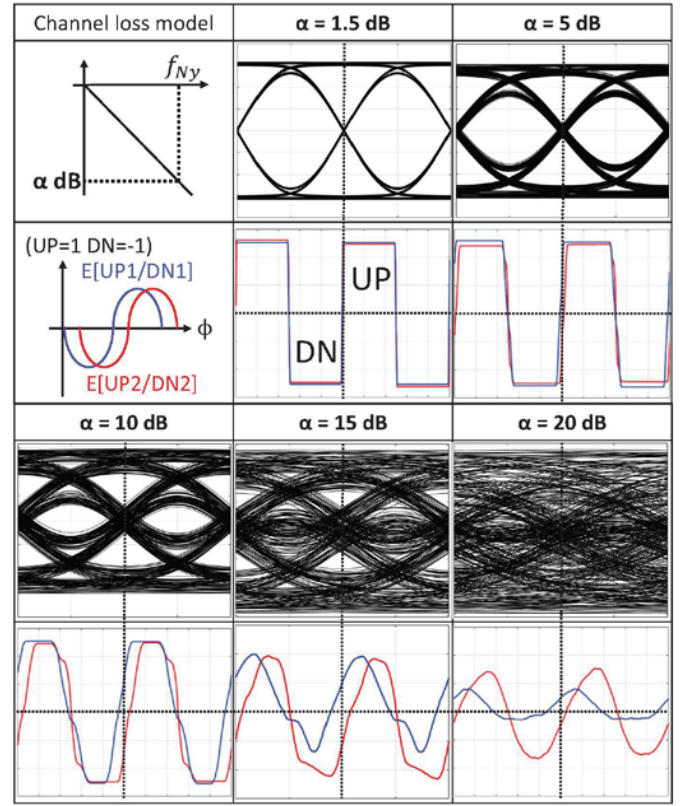

Fig. 4. Simulation results of two BBPD output groups classified by the exclusive-OR of two data bits

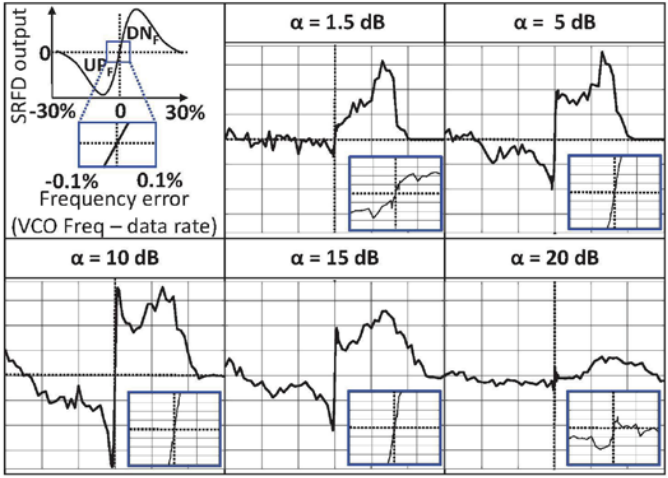

Fig. 5. simulation results of dispersion-based SRFD algorithm

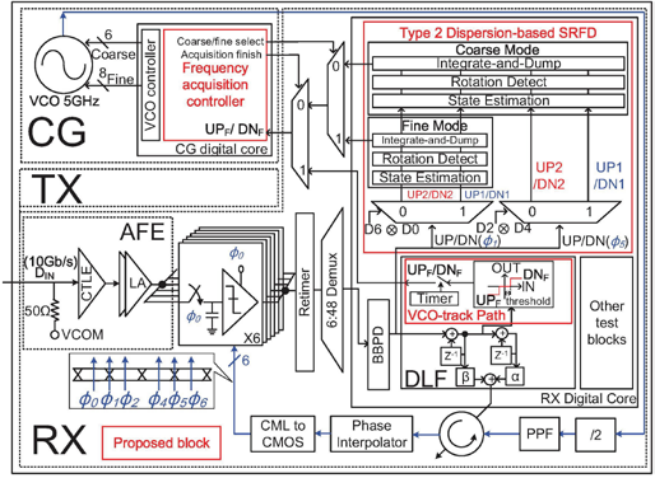

Fig. 6. The block diagram of 10Gbps reference-less transceiver

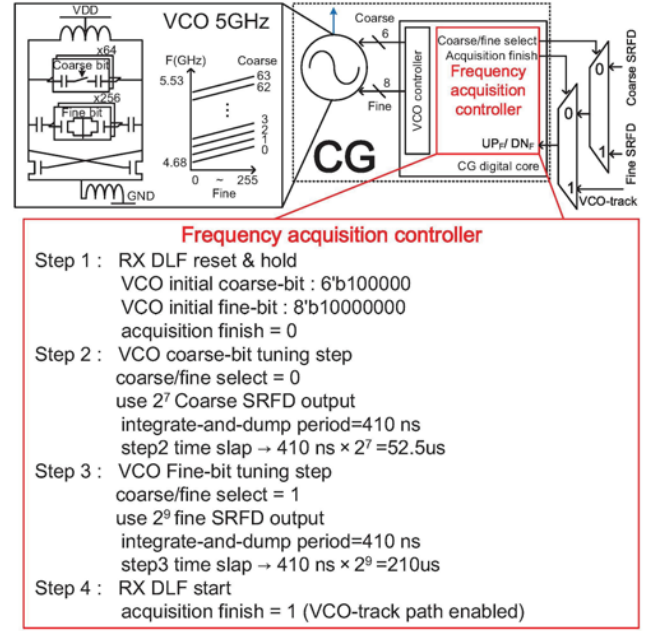

Fig. 7. The block diagram of Clock generator and frequency acquisition steps

## III. TRANSCEIVER ARCHITECTURE

Fig.6 illustrates a 10Gbps reference-less transceiver architecture featuring the proposed SRFD algorithm. The RX consists of an analog front-end (AFE), six samplers for quadrate operation, a re-timer, a demux, a phase rotator, and a synthesized digital core operating at a speed of 312.5 MHz. The digital core includes a variety of test algorithms for the type 2 dispersion-based SRFD and a digital loop filter(DLF). The dispersion-based SRFD operates in two modes; a coarse-mode using two edges sampled at $\phi_1$ and $\phi_5$ has a wide capture range, and a fine-mode uses only one edge sampled at $\phi_1$ to remove the frequency offset due to phase mismatch between $\phi_1$ and $\phi_5$.

Fig.7 summarizes the frequency acquisition procedure of the proposed scheme. The LC-VCO operating at 5GHz employs a six-bit MIM capacitor array and an eight-bit MOS capacitor array for coarse and fine frequency control, respectively. The operation sequence of the frequency acquisition is as follows. Firstly, the DLF and the control bits of the VCO are initialized. The VCO frequency is set to a center frequency to maximize the frequency detection gain of the SRFD algorithm within the VCO tuning range. Secondly, the VCO frequency is coarsely tuned to the input data rate by using $2^7$ $UP_F/DN_F$ results from the coarse-mode SRFD. Thirdly, the $2^9$ $UP_F/DN_F$ results from the fine-mode SRFD adjust the VCO frequency to match the input data rate precisely. Fourthly, the DLF is initiated for the phase lock. After recovering the input phase, the input data needs to be fully equalized for the optimum BER performance. Hence, an alternative feedback path instead of the dispersion-based SRFD is required to keep the VCO frequency staying within the pull-in-range under temperature and voltage variations. Even though the $2^{nd}$-order DLF tracks the frequency offset between the input data rate and the sampling clock, the stepwise operation of a rotator-based CDR degrades jitter and BER performance under large frequency offset. Therefore, in case the input value of the $2^{nd}$-order path exceeds the predefined threshold, the VCO-track path is activated to directly control the VCO frequency by generating $UP_F/DN_F$ signal every 100us for the compensation of the slow frequency drift without deteriorating the phase lock of the CDR. Finally, the continuous time linear equalizer (CTLE) is enabled to eliminate ISI for the optimum BER performance. The frequency lock time is 262.5us regardless of the initial frequency error.

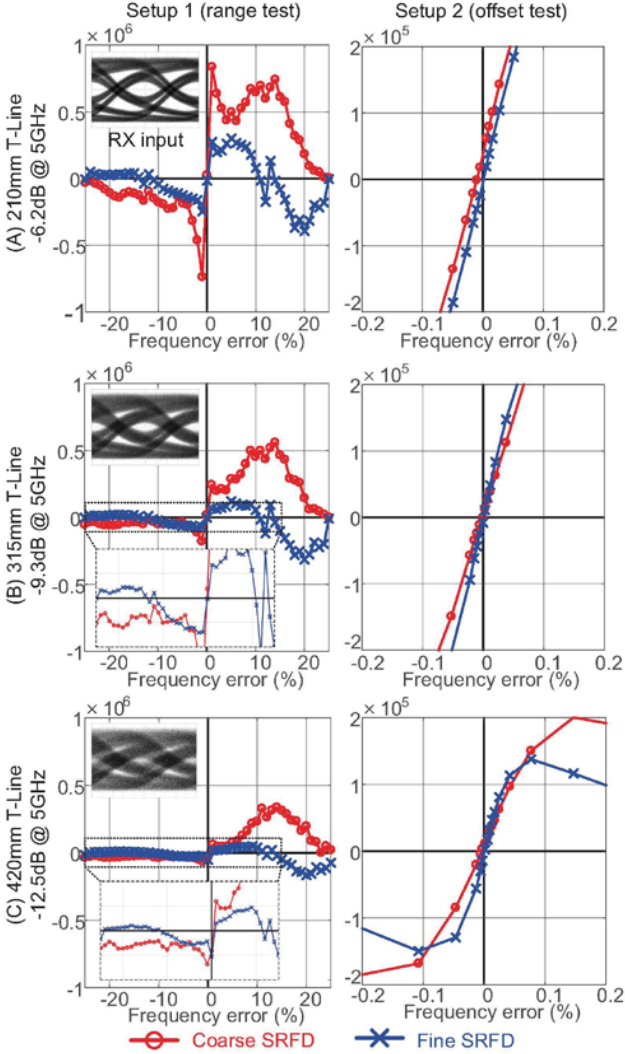

Fig. 8. Measured open loop output of type 2 dispersion-based SRFD

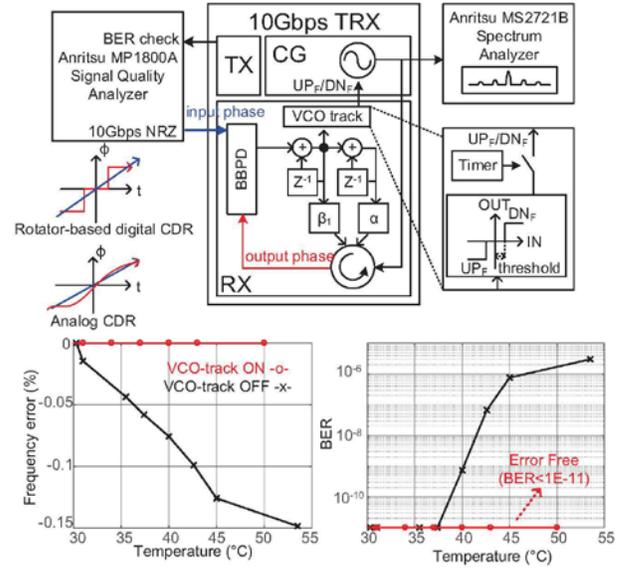

Fig. 9. RX clock recovered by the proposed frequency acquisition procedure

Fig. 10. Measured BER and VCO frequency versus chip temperature with and without VCO-track path

## IV. MEASUREMENT RESULTS

Fig.8 shows the open loop output of the dispersion-based SRFD tested using a PRBS31 NRZ. The period of the integrate-and-dump filter is extended to 13.4ms for accurate open loop output measurements. The first setup measures the SRFD output over 25% frequency range by sweeping the input data rate from 7.5Gbps to 12.5Gbps considering the limited tuning range of the 5GHz VCO. The second setup measures the exact frequency offset of the SRFD algorithm by sampling 10Gbps input data with a VCO frequency from 4.99GHz to 5.01GHz. Each setup is tested under three different channel conditions (A, B and C). A CTLE in an AFE is disabled to examine the SRFD algorithm under various ISI conditions. The post-layout simulation for the AFE shows the 3dB bandwidth of 6.7GHz, implying that the internal circuitry generates additional ISI. Therefore, the internal eye diagram is expected to be more dispersed than that shown in Fig.8. However, the dispersion-based SRFD accurately identifies the sign of the frequency error in three channel conditions. The frequency detection range of a coarse-mode SRFD is from −22% to 22%, which is 2x wider than that of the fine-mode SRFD.

The measured RX clocks recovered by the proposed frequency acquisition sequence at various data rates with 210mm PCB T-Line (channel A) are shown in Fig.9. The frequency acquisition scheme using the dispersion-based SRFD successfully recovers the input data phase within the maximum VCO tuning range from 4.7GHz to 5.6GHz.

Fig.10 shows the block diagram and test results of the VCO-track path that compensates for the frequency drift of the VCO due to temperature variations. The temperature of the test-chip is adjusted between 29°C and 53°C to validate the operation of the VCO-track path. The VCO without feedback from VCO-track path shifts the frequency by -1500ppm at 53°C and results in a significant increase in BER. However, when the VCO-track path is turned on, the VCO frequency remains constant without any bit-error detected.

The RX digital core consists of a BBPD, a DLF, an SRFD, and other diverse logic blocks for the test. The measured power consumption of the RX digital core with a retimer and a demux is 34.1mW under 1.2V supply for 10Gbps data-rate. Since the power consumption of the retimer and the demux is 9.6mW according to the post-layout simulation, the expected power consumption of the RX digital core is 24.5mW. The power and the area of the dispersion-based SRFD algorithm estimated by the equivalent gate count ratio are 9.3mW and 48900um2, respectively. The SRFD transceiver including the clock generator (CG) consumes 193.2mW. The chip photo fabricated in 65nm CMOS is shown in Fig.11.

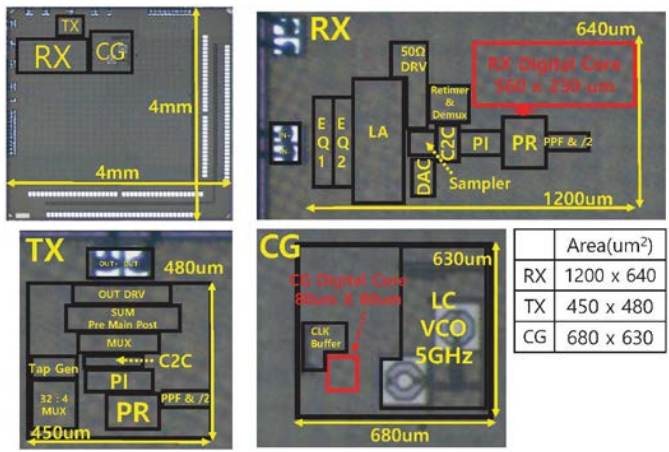

Fig. 11. Chip photo

V. CONCLUSION

This paper introduced a new frequency acquisition method using the SRFD algorithm for reference-less CDRs. SRFD estimates the phase state of input data using BBPD output with two different phase criteria and detects frequency error between input data and the sampling clock. After the frequency acquisition step, the VCO-track path in DLF keeps the VCO frequency within locking range of the CDR regardless of temperature or voltage variation. All of these algorithms can be implemented as synthesized digital blocks without additional analog circuitry, having significantly lower design efforts than other reference-less CDR design based on analog circuits. The 10Gbps transceiver including proposed algorithms is implemented in a 65nm CMOS process. Measured results demonstrate the feasibility of the proposed algorithm with detection range from −22% to 22%.